\documentstyle[twoside,fleqn,espcrc2,epsf]{article}
\epsfverbosetrue
\title{\vspace{-.5in}\hbox{}\hfill{\normalsize ANL-HEP-CP-97-74}\\
\vspace{.3cm}
Topology, fermionic zero modes and flavor singlet correlators in finite 
temperature QCD.}
 
\author{J.B. Kogut\address {Department of Physics, University of Illinois,
                            1110 West Green Street, Urbana IL 61801, USA}%
\thanks{Supported in part by NSF grant NSF-PHY92-00148.},
J.-F. Laga\"e\address {HEP Division, Argonne National Laboratory, 
                       9700 South Cass Avenue, Argonne IL 60439, USA}%
\thanks{Supported by DOE contract W-31-109-ENG-38.} 
and D.K. Sinclair$\null^{{\rm b}\dagger}$}

\begin{document}
 
\begin{abstract}
We extend our earlier results concerning the breaking of the $U_A(1)$ 
symmetry in finite temperature QCD. In particular, we use topological 
argments to investigate the chiral limit.
\end{abstract}
 

\maketitle

Among the non-perturbative properties of quarks, those associated with the 
topology of the gauge fields occupy a special place. This follows from the 
Atiyah-Singer index theorem which guarantees the existence of fermionic zero 
modes on configurations with non-trivial topology. As a result, 
non-perturbative issues related to the axial anomaly can generally be studied 
with enhanced control. 
In the context of finite temperature QCD, one such issue 
relates to the realization of the anomalous $U(1)$ axial symmetry. If this 
symmetry were to be effectively restored at the critical temperature, the 
increased number of soft degrees of freedom would modify the nature of the 
phase transition \cite{PISAR}. Needless to say, only the lattice can really 
hope to tackle this question, at this time.
Unfortunately, the lattice formulation also introduces artifacts and there 
is for example no exact index theorem in this context. Instead, what is 
observed with the actions currently used is the ``zero mode shift'' 
phenomenon \cite{VINK} (i.e. modes with small but non-zero eigenvalue and 
which are not exactly chiral). What we would like to show here is that even 
in this case, enough is recognizable of the continuum behavior that 
meaningful conclusions can be drawn. We find that the $U_A(1)$ 
symmetry is not restored at the phase transition but only at somewhat higher 
temperatures.  
At the same time, it is clear that detailed quantitative 
questions could only be answered with the use of an improved action.

Some of the results presented here are an extension of preliminary data 
introduced last year \cite{KLS}. Many of the trends are confirmed. 
In addition, we discuss the issue of reaching the chiral limit. 
We have measured the screening correlators in 
the scalar and pseudoscalar flavor singlet and flavor triplet sectors. 
From there, one can investigate the symmetry restorations by looking for 
degeneracies, either in the fitted screening masses or in the susceptibilities 
(integrated correlators). These computations were carried out on a set of 
configurations from the HTMCGC collaboration \cite{00625}. 
These were generated on a 
lattice of size $16^3\times8$ with 2 flavors of dynamical staggered fermions 
with mass $ma=0.00625$. In addition to those, we have generated a new set of 
configurations at $\beta =5.4875$ which is just above the crossover 
temperature at this value of the quark mass.
Our results for the screening masses are presented in figure 1. The new 
measurements at $\beta=5.4875$ confirm the trend observed earlier, namely 
that the $\sigma$ becomes light close to the transition while the $\delta$ 
does not. It is worth mentioning here that some of the fits presented in 
fig.1 may have large systematic errors (only the statistical errors are 
included in the figure). This is due in part to the small extent of the 
lattice and, in the case of the $\sigma$, to the additional difficulties 
associated with the measurements of disconnected quark loop correlators. 
Nevertheless, it seems to us that the trends observed in fig. 1 are rather 
clear. In addition, they are consistent with our measurements of the 
susceptibilities (as well as those by other groups \cite{SUSC}), 
in that the peak of the scalar susceptibility is seen in the disconnected 
and not the connected contribution (for summary plots of earlier results 
see ref \cite{Nt12}). If this 
situtation persists in the chiral limit, it will imply a non-restoration 
of the $U_A(1)$ symmetry in the high temperature phase of QCD. Namely, the 
$\pi$ and $\sigma$ become light close to the transition, but not the 
$\delta$. 

\begin{figure}[t]
%\vspace{-3.4cm}
\epsfxsize=7.5cm
\epsffile[70 70 550 550]{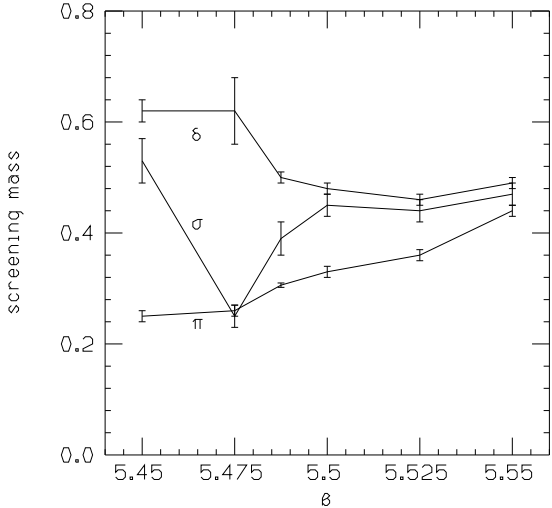}
\vspace{-1.2cm}
\caption{ Screening masses in two flavor QCD }
\vspace{-0.0cm}
\end{figure}

In the case of the susceptibilities, extrapolations to the chiral limit at 
fixed $\beta$ have been attempted by two groups but the results appear 
ambiguous \cite{CHRIST,DETAR}.
For example, it is seen that linear fits generally have a 
better $\chi^2$ than fits to a quadratic curve which would be expected on 
theoretical grounds. Here we would like to argue that these difficulties 
are due to lattice artifacts and will propose a method which allows us to 
extract information about the chiral limit without carrying out an explicit 
extrapolation. 
The idea is to identify the reason for the non-vanishing of the disconnected 
susceptibility in the chiral limit. This is done by making use of the spectral 
decomposition of the quark propagator and the resulting formulas:
\begin{eqnarray}
S(x,y) & = \sum_{\lambda} { \psi_{\lambda}(x) \psi^{\dagger}_{\lambda}(y) 
                   \over  i \lambda + m } \hfill \\
Q \equiv Tr S & = \sum_{\lambda > 0} { 2 m \over \lambda^2 + m^2 }
                  + { n_L + n_R \over m } \hfill \\
Q_5 \equiv Tr \gamma_5 S & = { n_L - n_R \over m } \hfill
\end{eqnarray}
where $n_L$ ($n_R$) are respectively the number of left (right) zero-modes 
of the Dirac operator. In terms of these the scalar and pseudoscalar 
disconnected susceptibilities are written 
$ \chi^{dis} = [<Q^2>-(<Q>)^2]/V $ and $ \chi^{dis}_5 = <Q^2_5>/V $.
Now it is clear that only the contribution from fermionic zero-modes will 
survive at $m=0$ and that the chiral limit will be smooth. 
Remember that for $N_f=2$ the fermionic determinant picks up a factor of 
$ m^2 $ for each zero mode. On configurations with topological charge one, 
the factors of m will cancel from measurements of $ \chi^{dis}_{(5)} $ and 
leave a well defined non-zero result.
This is what we expect in the continuum, on the lattice however things will 
be more complicated. First, there are no exact zero modes ( except on a 
subspace of measure zero ). Then:
\begin{eqnarray}
Q^{Latt} & = \sum_{\lambda > 0} { 2m \over \lambda^2 + m^2 } \hfill \\
Q^{Latt}_5 & = \sum_{\lambda > 0} { 2m < \psi_{\lambda} | \Gamma_5 | 
             \psi_{\lambda} > \over \lambda^2 + m^2 } \hfill
\end{eqnarray}
where we have used the symmetries of the staggered action and $\Gamma_5$ is 
the (four-link) lattice $\gamma_5$ operator.
It is then clear that on the lattice, disconnected susceptibilities will 
{\it vanish} in the chiral limit and be proportional to $m^2$ for small $m$. 
We believe that it is this lattice artifact which makes a chiral extrapolation 
from the lattice data extremely difficult \cite{CHRIST,DETAR}.

\begin{figure}[htb]
%\vspace{-3.4cm}
\epsfxsize=7.5cm
\epsffile[70 70 550 550]{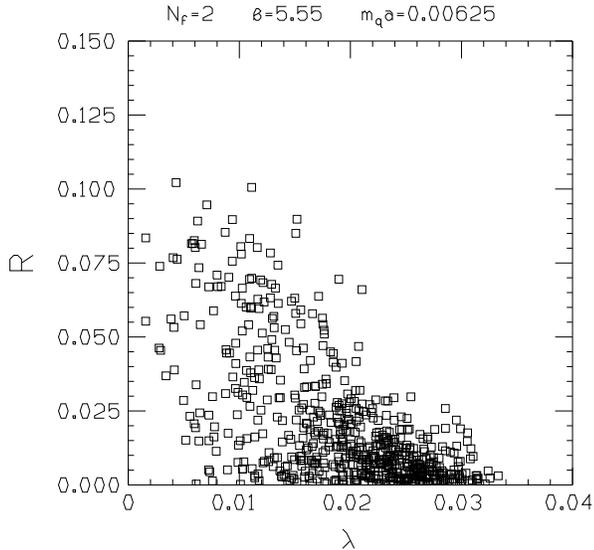}
\vspace{-1.2cm}
\caption{ $(r,\lambda)$ diagram at $\beta=5.55$ }
\vspace{-0.2cm}
\end{figure}

\begin{figure}[htb]
%\vspace{-3.4cm}
\epsfxsize=7.5cm
\epsffile[70 70 550 550]{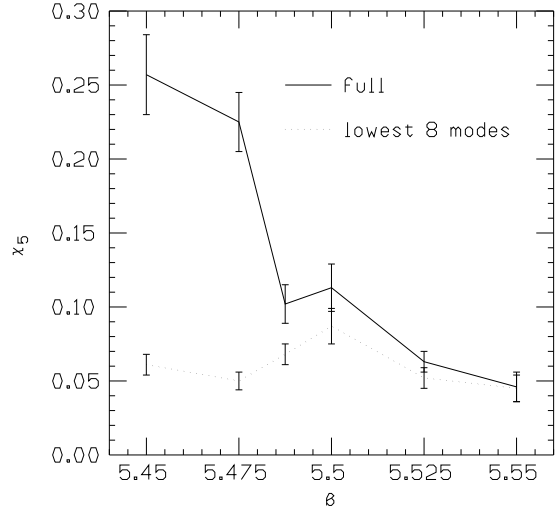}
\vspace{-1.2cm}
\caption{ disconnected $\gamma_5$ susceptibility }
\vspace{-0.2cm}
\end{figure}

We therefore propose a different method for studying the 
$m \rightarrow 0$ limit. The idea is to identify ``continuum behavior'' 
in the lattice data (and then use what we know about the continuum theory). 
Together with the Atiyah-Singer index theorem, (3) implies $m Q_5 = Q_{top}$. 
We would like to see the same connection with topology on the lattice. 
The formula (3) comes about because in the continuum $<\psi_{\lambda}| 
\gamma_5|\psi_{\lambda}>$ is either $\pm 1$ if $\lambda=0$ or $0$ otherwise. 
The lattice expression (5) will closely match the continuum if we see on the 
lattice a clear correlation between small eigenvalues and large 
$r_{\lambda} \equiv <\psi_{\lambda}|\Gamma_5|\psi_{\lambda}>$ 
(i.e. if we can identify chiral modes). Figure 2 is a plot of 
$r_{\lambda}$ versus $\lambda$ for the lowest 8 positive 
eigenvalues on each configuration of our sample at $\beta=5.55$. Many of the 
small eigenvalues are indeed associated with a larger value of $r_{\lambda}$.
Of course, 
the quality of the correlation depends on how strong the coupling is ( It is 
for example not as clear here as it was in quenched QCD at $\beta=6.2$ 
\cite{KLS} ). 
The second step in the argument consists in showing that these low modes 
(the would be chiral modes of the continuum theory) indeed saturate the 
lattice results. In figure 3 we compare the full result for $\chi^{dis}_5$ 
with the one obtained by truncating (5) to the lowest 8 modes. The agreement 
is perfect at $\beta=5.55$ and deteriorates somewhat in the rest of the 
symmetric phase (in part because of lattice artifacts). Putting everything 
together, we find that $\chi^{dis}_5$ is non-zero in the symmetric phase and 
that the topological origin of this result guarantees that it will remain 
non-zero in the {\it continuum} chiral limit. We therefore conclude that the 
$U_A(1)$ symmetry is not restored at the phase transition 
\footnote{ Of course, this conclusion only concerns the volume 
studied so far, namely $V=(2/T)^3$. The actual volume dependance remains 
to be investigated. }. Our results
also indicate the importance of investigating improved lattice actions 
which better satisfy the index theorem. These would allow the continuum 
behavior to appear even more clearly than here. Also, lower values of the 
quark mass would be allowed before the zero-mode shift phenomenon becomes 
a problem. Certainly, such actions will be necessary to obtain 
quantitative results. In fact, some of the formulations that one would like 
to consider have been mentioned at this conference (domain wall fermions, 
perfect action, ...). Beyond that, there are many other questions which 
can be investigated with the knowledge of the low lying eigenvalues and 
eigenvectors. Correlations between the eigenvalues is one of them. 
The relation with instantons and the possible existence of instanton- 
anti-instanton molecules is another.


\begin{thebibliography}{99}
\bibitem{PISAR} R.D.Pisarski and F.Wilczek,Phys.Rev.D{\bf 29}, 338(1984)
\bibitem{VINK} J.C.Vink, Phys. Lett. {\bf 210}B (1988) 211
\bibitem{KLS} J.B.Kogut, J.-F.Laga\"e and D.K.Sinclair,Nucl. Phys.B(PS)
{\bf53} (1997)269
\bibitem{00625} S.Gottlieb et al., Phys.Rev.D{\bf 55},6852(1997)
\bibitem{SUSC} F.Karsch and E.Laermann,Phys.Rev.D{\bf 50}, 6954(1994);
C.Bernard et al., Phys. Rev. D{\bf 45},3854(1992)
\bibitem{Nt12} C.Bernard et al., Phys.Rev. D{\bf 54},4585 (1996)
\bibitem{CHRIST} N.Christ, Nucl.Phys. B(PS){\bf 53} (1997)253
\bibitem{DETAR} C.Bernard et al., Phys.Rev.Lett.{\bf 78} (1997)598
\end{thebibliography}
\end{document}